\documentclass[conference]{IEEEtran}
\usepackage{caption}
\usepackage[none]{hyphenat}
\usepackage{algpseudocode}
\usepackage{amsmath}
\usepackage{amsfonts}
\usepackage{amssymb}
\usepackage{subfigure}
\usepackage{graphicx}
\usepackage[table]{xcolor}
\usepackage{array,multirow}

\usepackage{amsfonts}
\usepackage{booktabs}
\usepackage{siunitx}

\usepackage{color}

\usepackage{tikz} 
\newcommand*\circled[1]{\tikz[baseline=(char.base)]{%
        \node[shape=rectangle,fill=gray!20,draw,inner sep=2pt,opacity=0.5,text opacity=1] (char) {#1};}}

\newcommand\citem{%
  \stepcounter{enumi}\item[\circled{\arabic{enumi}}]}

\usepackage{enumitem}

\usepackage{url}
\makeatletter
\def\url@leostyle{%
  \@ifundefined{selectfont}{\def\UrlFont{\sf}}{\def\UrlFont{\small\ttfamily}}}
\makeatother
\newcommand{\eat}[1]{}
\usepackage[linesnumbered,ruled]{algorithm2e}
\usepackage{mdframed} 

\renewcommand{\IEEEbibitemsep}{1pt plus 2pt}
\makeatletter
\IEEEtriggercmd{\reset@font\normalfont\small}
\makeatother
\IEEEtriggeratref{1}

\usepackage{xcolor}
\definecolor{light-gray}{gray}{0.9}

\newenvironment{packed_enum}{%
  \begin{enumerate}%
  }{\end{enumerate}}

\usepackage{amsthm}

\newtheorem{lemma}{Lemma}

\newcolumntype{L}[1]{>{\raggedright\let\newline\\\arraybackslash\hspace{0pt}}m{#1}}

\begin{document}


\title{Decentralized Privacy-preserving Timed Execution in Blockchain-based Smart Contract Platforms}

\eat{
\author{%
{Chao Li and Balaji Palanisamy }%
\fontsize{10}{10}\selectfont\itshape
School of Computing and Information, University of Pittsburgh, USA\\
\{chl205, bpalan\}@pitt.edu\\
\fontsize{10}{10}\selectfont\rmfamily\itshape
}%
}
\author{

\IEEEauthorblockN{Chao Li}
\IEEEauthorblockA{School of Computing and Information\\
University of Pittsburgh\\
Pittsburgh, USA\\
Email: chl205@pitt.edu}

\and

\IEEEauthorblockN{Balaji Palanisamy}
\IEEEauthorblockA{School of Computing and Information\\
University of Pittsburgh\\
Pittsburgh, USA\\
Email: bpalan@pitt.edu}

}

\maketitle

\begin{abstract}
Timed transaction execution is critical for various decentralized privacy-preserving applications powered by blockchain-based smart contract platforms. Such privacy-preserving smart contract applications need to be able to securely maintain users' sensitive inputs off the blockchain until a prescribed execution time and then automatically make the inputs available to enable on-chain execution of the target function at the execution time, even if the user goes offline. While straight-forward centralized approaches provide a basic solution to the problem, unfortunately they are limited to a single point of trust. This paper presents a new decentralized privacy-preserving transaction scheduling approach that allows users of Ethereum-based decentralized applications to schedule transactions without revealing sensitive inputs before an execution time window selected by the users. The proposed approach involves no centralized party and allows users to go offline at their discretion after scheduling a transaction. The sensitive inputs are privately maintained by a set of trustees randomly selected from the network enabling the inputs to be revealed only at the execution time. The proposed protocol employs secret key sharing and layered encryption techniques and economic deterrence models to securely protect the sensitive information against possible attacks including some trustees destroying the sensitive information or secretly releasing the sensitive information prior to the execution time. We demonstrate the attack-resilience of the proposed approach through rigorous analysis. Our implementation and experimental evaluation on the Ethereum official test network demonstrates that the proposed approach is effective and has a low gas cost and time overhead associated with it.   


\end{abstract}

\section{Introduction}
\label{s1}

In the age of big data, blockchain~\cite{nakamoto2008bitcoin} has become a promising technology to enable decentralized protection of data integrity~\cite{gaetani2017blockchain} and ensuring data quality~\cite{azaria2016medrec}.
Any data stored in a blockchain is backed up and verified by all the nodes in the network and provides a strong resilience against attacks that can tamper the integrity of the data.
With this great feature offered by blockchains, recent implementations of blockchain-based smart contract platforms, such as Ethereum~\cite{wood2014ethereum} and NEO~\cite{NEO}, have attracted a large number of developers to build decentralized applications using smart contracts that avoid the need of a centralized server to manage and maintain the data \cite{Auctionhouse,mccorry2017smart,miller2017zero}. 
The market cap for the leading smart contract platform, Ethereum, peaked at \$134 billion~\cite{ETH} in 2018 and thousands of decentralized applications, ranging from social networks to financial software, have been developed over Ethereum~\cite{DAPP}.
The Smart Contracts market is estimated to grow at a CAGR of 32\% during the period 2017 to 2023~\cite{marketresearchfuture}.

A decentralized application may involve one or more smart contracts and each smart contract may contain multiple functions that need to be invoked by application users through transactions.
For instance, a sealed-bid auction smart contract~\cite{SealedBidAuction} requires bidders to reveal their sealed bids by invoking a function (e.g., a \textit{reveal() function}) during a time window. Similarly, a voting smart contract~\cite{mccorry2017smart} requires voters to publish their votes using a \textit{vote()} function during the voting time window.
Each called function in a smart contract is executed by the entire blockchain network.
Since both function code and function inputs (i.e., bid or vote) are available on the blockchain, the function outputs are deterministic and their correctness can be verified by the network, 
thus cutting out centralized middlemen or intermediaries for running these functions~\cite{lauslahti2017smart}.


A key fundamental limitation of existing smart contract platforms is the lack of support for users to schedule timed execution of transactions such that their target functions can be invoked at a later time, even when the users go offline.
For example, if Bob plans to take a week off work and could not respond to an auction or voting mechanism implemented on Ethereum during the prescribed time windows, he needs a mechanism to schedule these timed transactions by automatically invoking \textit{reveal()} and \textit{vote()} during the time windows. Here, the inputs to these functions namely the bids and the votes are extremely sensitive and need to be securely protected until the prescribed time windows even when Bob is offline.
Scheduling timed execution of functions is common in centralized application environments.
For instance, Boomerang~\cite{Boomerang} allows users of Gmail to schedule their emails to be sent when users have no connection with the Internet. 
Similarly, Postfity~\cite{Postfity} helps users to schedule messages to be posted onto many centralized social networks.
However, a centralized approach to supporting timed execution of transactions incurs a single point of trust and violates the key design principle of decentralization inherent in blockchain-based smart contract platforms~\cite{ins1}.
In general, the design of timed execution of transactions in decentralized platforms such as Ethereum is challenged in two aspects. First, when a transaction invoking a function is deployed into the network, the invoked function is executed immediately which makes it difficult to support timed execution when the user has already gone offline. Another key challenge arises due to privacy concerns associated with the input data to the function. To guarantee verifiability of function outputs, function inputs need be put onto the blockchain and as a result, both function inputs and outputs become public to all peers at the time the schedule is initialized leading to privacy risks with the input data. 
The proposed privacy-preserving timed-execution approaches find numerous applications in high performance computing.
For instance, recent projects such as Golem~\cite{Golem} and iEx.ec~\cite{iEx.ec} focus on developing decentralized supercomputers and high performance computing platforms without vendor lock-in. These solutions leverage the Ethereum as a marketplace application to link buyers and sellers of computing resources without requiring an intermediary. When smart contracts are used to manage and schedule computing tasks in such platforms, privacy-preserving timed-execution techniques can effectively protect privacy of sensitive inputs of the scheduled and in-queue computing tasks.

In this paper, we design and develop a new decentralized privacy-preserving timed execution mechanism that allows users of Ethereum-based decentralized platforms to schedule timed execution of transactions without revealing function inputs and outputs prior to the execution time selected by the users. 
The proposed approach is decentralized and involves no centralized party and does not include any single point of trust. After transactions have been scheduled, it requires no further interaction from users and allows users to go offline at their discretion.
The mechanism does not reveal function inputs before the execution time window selected by a user as function inputs are privately maintained by a set of trustees randomly selected from the network and released only during the execution time window.
The function inputs are protected through secret share~\cite{shamir1979share} and multi-layer encryption~\cite{dingledine2004torr} and possible misbehaviors of the trustees are made detectable and verifiable through a suit of misbehavior report mechanisms implemented in the Ethereum Smart Contracts and any verified misbehavior incurs monetary penalty on the violator.
We implement the proposed approach using the contract-oriented programming language \textit{Solidity}~\cite{Solidity2017} and test it on the Ethereum official test network \textit{rinkeby}~\cite{Rinkeby2017} with Ethereum official Go implementation \textit{Geth}~\cite{Geth2017}. 
Our implementation and experimental evaluation that the proposed approach is effective and the protocol has a low gas cost and time overhead associated with it.

\section{Overview of timed execution in Ethereum}
\label{s2}

In this section, 
we first present the preliminaries of the Ethereum smart contract platform~\cite{wood2014ethereum} and describe the challenges involved in implementing timed execution of smart contracts over Ethereum. 
We then present the key ideas behind the proposed solution and introduce the organization of the proposed protocol and discuss the security challenges and potential attacks encountered in the proposed approach.

\subsection{Preliminaries}

\label{s2.0}

A blockchain represents a decentralized and distributed public digital ledger that guarantees that the records stored in it cannot be tampered without compromising a majority of nodes in the network~\cite{nakamoto2008bitcoin}.
Then, a smart contract is a piece of program code stored in a blockchain and it usually consists of multiple functions.
In the leading smart contract platform Ethereum~\cite{wood2014ethereum}, there are two types of accounts, namely External Owned Accounts (EOAs) controlled by private keys and Contract Accounts (CAs) for storing smart contract code.
An Ethereum node can create as many as EOAs and then use EOAs to create CAs by deploying smart contracts. 
However, since smart contracts are passive, their execution must be invoked through either a transaction sent by an EOA or a message sent from a CA.
As a result, the transactions/messages, as well as function inputs inside them, are all recorded by the Ethereum blockchain, which makes the function outputs deterministic because all Ethereum nodes can execute the function with the same inputs and gets the same outputs.
In Ethereum, to deploy a smart contract (i.e., CA) or call a smart contract function changing any data on blockchain, one needs to pay for Gas~\cite{wood2014ethereum}. Gas can be exchanged with Ether, the cryptocurrency used in Ethereum, and Ether can be exchanged with real money. 


\subsection{Problem statement}

\label{s2.1}
\begin{figure}
\centering
{
    \includegraphics[width=9cm,height=4cm]{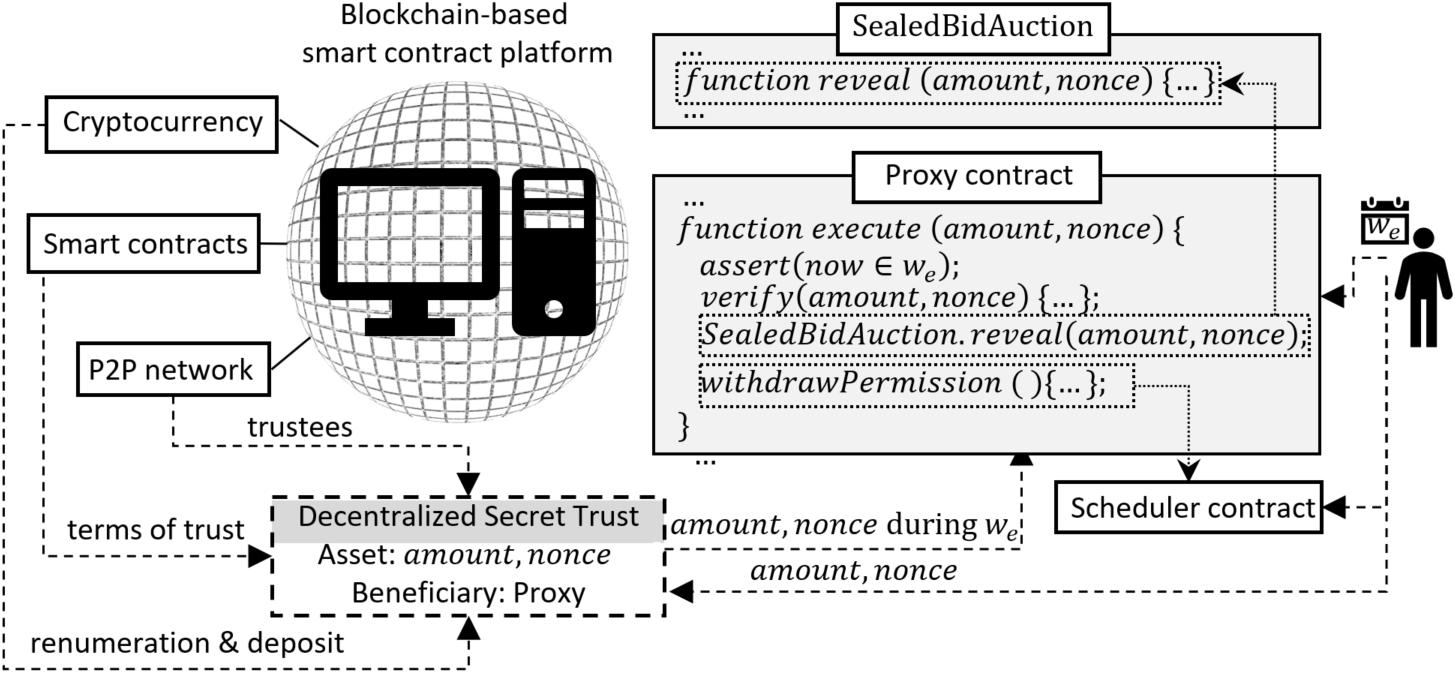}
}
\caption {\small At time $t_1$, Bob wants to schedule function \textit{reveal(amount,nonce)} in contract \textit{SealedBidAuction}~\cite{SealedBidAuction} to be executed during a future time window $w_e$}
\label{f3} 
\end{figure}

The Ethereum blockchain platform~\cite{wood2014ethereum} can be viewed as a giant global computer as shown in Figure~\ref{f3}.
If a user creates a EOA and uses the EOA to send a transaction with inputs $x_1$ and $x_2$ to call function $f(x_1,x_2)$ of a smart contract $C$ at time $t_1$, 
function $f(x_1,x_2)$ will be executed instantly and the inputs $x_1$ and $x_2$ will be made public. 
This is acceptable if the user just wants to reveal $x_1$ and $x_2$ at time $t_1$.
However, if the user needs to reveal $x_1$ and $x_2$ during a future execution time window $w_e$, sending the transaction at $t_1$ will not work.
For example, Bob may want to make function $reveal(amount,nonce)$ of a sealed-bid auction smart contract~\cite{SealedBidAuction} be executed during a future execution time window $w_e$. Then, sending the transaction out at $t_1$ will make his bid value be known to all other bidders immediately, which violates his privacy requirements.

\subsection{Privacy-preserving timed execution}

\label{s2.2}
To support privacy-preserving timed execution of smart contracts, the transaction calling function $f(x_1,x_2)$ must be sent during the prescribed execution time window $w_e$ while inputs $x_1$ and $x_2$ should not be revealed before $w_e$.
Our proposed protocol for supporting privacy-preserving timed execution is implemented as two smart contracts, namely a unique scheduler contract $C_s$ managing all schedule requests of users in Ethereum and a proxy contract $C_p$ deployed by each user having a schedule request.
At the time of setting a timed execution, the protocol requires the user to (1) store schedule information, including a cryptographic \textit{keccak-256} hash \cite{bertoni2011keccak} of function inputs $x_1$ and $x_2$ to the scheduler contract $C_s$, (2) deploy a proxy contract $C_p$ and (3) employ a group of EOAs as trustees.
The main functionality of the proxy contract $C_p$ is implemented through a function $execute()$ in it. 
Once $C_p$ receives a transaction during $w_e$ with the desired inputs $x_1$ and $x_2$ verified through their hashes in scheduler contract $C_s$, the function $execute()$ will immediately send a message calling the target function $f(x_1,x_2)$ with inputs $x_1$ and $x_2$.
The trustees are in charge of storing inputs $x_1$ and $x_2$ off the blockchain before the execution time window $w_e$ and they send a transaction with the inputs to the proxy contract $C_p$ during $w_e$. 
The terms of the decentralized secret trust created by the user as a settlor, namely what the trustees can or cannot do, are programmed as functions in smart contracts $C_s$ and $C_p$.
Each trustee needs to pay a security deposit $d$ (i.e., Ether) to the scheduler contract $C_s$ and any detectable misbehavior of this trustee makes the deposit be confiscated. The security deposit serves as an economic deterrence model for enforcing behaviors of peers in the blockchain network~\cite{andrychowicz2014secure,miller2017zero}.
Finally, after the trustees have sent a transaction with inputs $x_1$ and $x_2$ to the proxy contract $C_p$ during $w_e$, they can withdraw both their deposit and remuneration paid by the user from the scheduler contract $C_s$.
In the example of Figure~\ref{f3}, at $t_1$, Bob stores hash of inputs $amount$ and $nonce$ to $C_s$, deploys $C_p$ and employs a group of trustees. 
These trustees, after signing an agreement with Bob, are in charge of revealing the asset $amount$ and $nonce$ to the beneficiary, proxy contract $C_p$, during $w_e$. During the execution time window $w_e$, after the trustees have sent a transaction with inputs $amount$ and $nonce$ to $C_p$, the function $execute()$ in $C_p$ can trigger $reveal()$ in the \textit{SealedBidAuction} contract through $SealedBidAuction.reveal()$ and also unlock trustees' deposit and remuneration in $C_s$ through $withdrawPermission()$.

\subsection{Protocol overview}

\label{s2.3}
\begin{figure}
\centering
{
    \includegraphics[width=8cm,height=4cm]{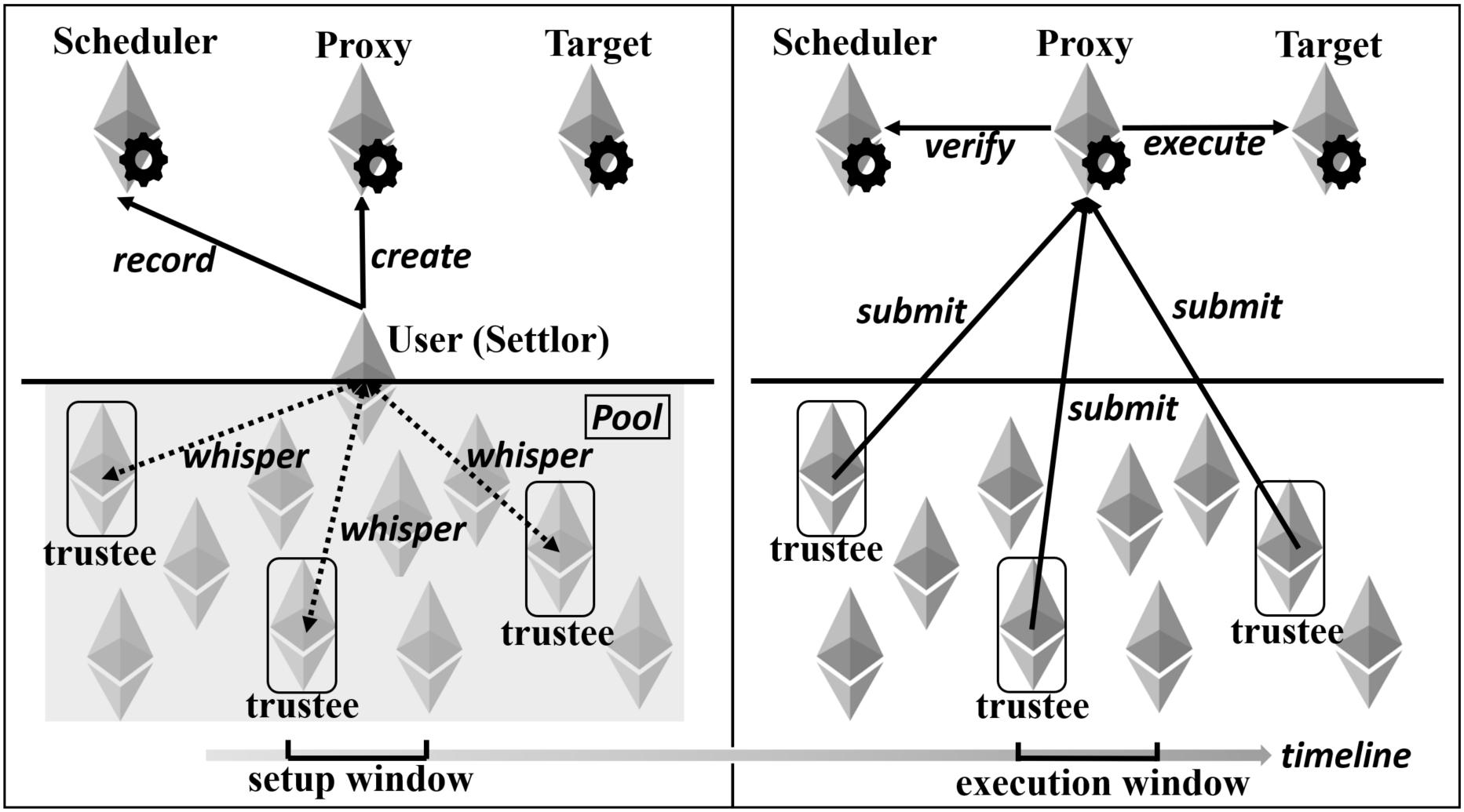}
}
\caption {Protocol overview}
\label{f1} 
\end{figure}

The proposed protocol consists of four components:

\noindent \textbf{\textit{Trustee application}}:
At any point in time, an EOA can apply to $C_s$ for getting added into a trustee candidate pool maintained by $C_s$ by submitting its working time window and paying a security deposit.  
During the working time window, the EOA should be able to connect with Ethereum to send transactions to the proxy contract $C_p$. 
In the example shown in Figure~\ref{f1}, we notice that ten EOAs joined the pool.
The public pool then makes the entire network learn that this EOA can provide services during its declared working times.

\noindent \textbf{\textit{User schedule}}:
During setup time window $w_s$, a user can schedule a transaction by registering the schedule to scheduler contract $C_s$, deploying a proxy contract $C_p$, and secretly selecting trustees from the pool. The selected trustees should keep the function inputs privately before the execution time window $w_e$ while revealing them during $w_e$ to make the target function be executed.
In Figure~\ref{f1}, during setup window $w_s$, the user informed the schedule with the scheduler contract $C_s$ and deployed the proxy contract $C_p$. Then, the user randomly selected three EOAs from the pool as trustees and signed agreements with the trustees through private channels created by the whisper protocol~\cite{Whisper2017}. 
Any data exchanged through the whisper channels are encrypted and can only be viewed by the data sender and data recipient.

\noindent \textbf{\textit{Function Execution}}:
During execution time window $w_e$, the selected trustees submit the function inputs to the proxy contract $C_p$ through transactions, which triggers $C_p$ to verify correctness of function inputs with $C_s$ and then call the scheduled function in the target contract $C_t$.
In Figure~\ref{f1}, during $w_e$, the trustees submitted stored data to proxy contract $C_p$. After verifying the received data with the hashes stored in scheduler contract $C_s$, $C_p$ called the function in $C_t$.

\noindent \textbf{\textit{Misbehavior report}}:
During the entire process, trustees may perform several types of misbehaviors violating the protocol, such as secretly disclosing stored data before $w_e$ or rejecting to submit stored data during $w_e$. To tackle these issues, the protocol involves several misbehavior report mechanisms that allow any witness of a misbehavior to report it to the scheduler contract $C_s$ and earn a component of the deposit paid by the suspect trustee once the report is verified to be true.

\subsection{Security challenges and attack models}

\label{s2.4}
The proposed mechanism encounters several critical security challenges, which can be roughly classified using two attack models.

\noindent \textbf{Time difference attacks}:
The time difference attack happens when an adversary aims at obtaining the function inputs at a time point $t_d$ earlier than the execution time window $w_e$ so that he can leverage the time difference between $t_d$ and $w_e$ to achieve his purpose. There are two key methods to launch a time difference attack.

\begin{itemize}[leftmargin=*]


\item \textbf{Trustee identity disclosure}:
In \textit{user schedule} component of the protocol, trustees are secretly selected by user $U$. Therefore, from the perspective of EOAs besides the selected trustees and user $U$, all EOAs in network with working time windows satisfying $U$'s requirement have equal chance to be selected by $U$, thus protecting the identifications of selected trustees with highest entropy and uncertainty. However, a trustee, after being selected, may announce its identity to the public to seek trade with potential adversaries about the stored data. To prevent such misbehavior, the proposed protocol employs a trustee identity disclosure report mechanism in \textit{misbehavior report} component of the protocol, which forces a trustee to disclose its identity with the sacrifice of the confiscation of its security deposit.

\item \textbf{Advance disclosure}:
A trustee may choose to voluntarily disclose the stored data to the entire network without seeking bribery. To penalize such misbehavior, an advance disclosure report mechanism is employed  in the \textit{misbehavior report} component, 
which makes any trustee disclosing its stored data in advance lose its security deposit.

\end{itemize}

\noindent \textbf{Execution failure attack}:
The execution failure attack happens when an adversary aims at making the execution of the target function fail during the execution time window $w_e$.
There are two key methods to launch this attack.

\begin{itemize}[leftmargin=*]

\item \textbf{Absent trustee}: 
A trustee may become absent during the execution time window $w_e$, which makes its stored data get lost. To prevent this type of misbehavior, the \textit{user schedule} component of protocol requires each selected trustee to provide a signature, which will only be revealed along with the function inputs during $w_e$. Therefore, before $w_e$, the identities of trustees are kept secret. In contrast, during $w_e$, the identities become public so that any present trustee can report an absent trustee through the absent trustee report mechanism in the \textit{misbehavior report} component of protocol, 
which penalizes any absent trustee by confiscating its security deposit.


\item \textbf{Fake submission}:
A trustee may submit fake stored data to the proxy contract $C_p$ during $w_e$, which may cause the restoration of the function inputs to fail. The protocol handles this type of misbehavior using the fake submission report mechanism in the \textit{misbehavior report} component of protocol, 
which confiscates violator's security deposit if its submission is proved to be fake.


\end{itemize}

\section{Protocol description}
\label{s3}
In this section, we present the proposed protocol organized along the four components introduced in Section~\ref{s2.3}.

\subsection{Trustee application}

The first component \textit{trustee application} allows EOAs that want to earn remuneration through the trustee job to register to the scheduler contract $C_s$ and make their information public.
There are three key steps in this component.
We note that a step with a gray bullet (e.g., \colorbox{gray!18}{1}) refers to an off-chain action not recorded by blockchain while a step with a white bullet (e.g., 2) refers to an on-chain action recorded by blockchain. We will distinguish off-chain and on-chain steps with the two bullet types in all four components of the protocol.

\begin{small}
\begin{mdframed}[innerleftmargin=8pt]
\centerline{\textbf{Trustee application}}
\textbf{Input:} scheduler contract $C_s$

\noindent \textbf{Apply:} 
\begin{packed_enum}[leftmargin=*]
  \citem An Ethereum node creates a new EOA.
  \item This EOA applies to the scheduler contract $C_s$ for being added into the trustee candidate pool by submitting a public key, a whisper key, working time window, a security deposit and a beneficiary address.
  \item The scheduler contract $C_s$ verifies the application and accept the application if all required data has been submitted.
\end{packed_enum}

\end{mdframed}
\end{small}

\noindent \textbf{Step 1}:
Each trustee candidate should be a newly generated EOA, which only has an amount of Ether (the cryptocurrency in Ethereum) that will be submitted to the scheduler contract $C_s$ as security deposit $d$ in step 2. No additional Ether should be left because we will need the account to make its account private key public during execution time window $w_e$.

\noindent \textbf{Step 2-3}:
An EOA should apply for the trustee candidate by sending a transaction to $C_s$ with the five listed information.

\begin{itemize}[leftmargin=*]

\item The public key will later be used by user $U$ in step 8 of \textit{user schedule} component to generate $onions$~\cite{dingledine2004torr}. Here, the term $onion$ refers to the output of iteratively encrypting data with multiple public keys.

\item The whisper key will later be used by user $U$ in \textit{user schedule} component to establish private channel with this EOA through whisper protocol~\cite{Whisper2017}. 

\item The working time window will be used by user $U$ in step 6 and 10 of \textit{user schedule} component to select trustees satisfying $U$'s requirements (i.e., execution time window). 

\item The security deposit is a fixed amount of Ether hard-coded in scheduler contract $C_s$. Once being submitted to $C_s$, the deposit can only be withdrawn at the end of EOA's working time window, if there is no misbehavior reported through report mechanisms in \textit{misbehavior report} components.

\item Finally, the protocol needs the EOA to make its account private key public in \textit{function execution} component, so the beneficiary address will be the address of a safe EOA to receive deposit and remuneration withdraw.

\end{itemize}


\subsection{User schedule}

The second component \textit{user schedule} prescribes how a user should set a schedule through  three key operations, namely deploying a proxy contract (step 3), registering the schedule information to scheduler contract $C_s$ (step 4) and implementing a two-round trustee selection (step 5-13).

For the illustration of the protocol in step 5 to 13, we will use the example shown in Figure~\ref{f2}.

\begin{small}
\begin{mdframed}[innerleftmargin=8pt]
\centerline{\textbf{User schedule}}
\textbf{Input:} scheduler contract $C_s$, target contract $C_t$

\noindent \textbf{Initialization:} 
\begin{packed_enum}[leftmargin=*]
  \citem User $U$ decides function inputs $IN$, execution time window $w_e$, secret sharing parameters $(m,n)$, number of layers $l$, a 256-bit secret key $key$ and a 256-bit random number $R_U$.
  \citem User $U$ computes the remuneration $r$.
  \item User $U$ deploys proxy contract $C_p$ to the network.
  \item User $U$ registers the schedule to scheduler contract $C_s$ with $(w_e,m,n,l,C_p^{addr},r)$ and receive a schedule ID $sid$.
  \citem User $U$ splits $key$ to $n$ $shares$ through $(m,n)$ secret sharing.
  
\end{packed_enum}

\noindent \textbf{First-round trustee selection:} 
\begin{packed_enum}[leftmargin=*]
  \setcounter{enumi}{5}
  \citem User $U$ randomly selects $n(l-1)$ trustees and sends each trustee a $(sid,tid)$, where $tid$ refers to a non-repeated ID in the range of $[0,n(l-1))$ assigned to the trustee.
  \citem Each selected trustee $T$ then does the following:
  \begin{packed_enum}[leftmargin=*]
  	\item Verify $(U^{addr},sid,tid,w_e,r)$ with $C_s$.
  	\item Generate a 256-bit random number $R_T$.
  	\item Take keccak256 hash $h(T^{addr},R_T)$.
  	\item Sign $(U^{addr},sid,tid,h(T^{addr},R_T))$ with $T$'s private key, which gives signature $vrs=(v,r,s)$.
  	\item Send $h(T^{addr},R_T)$ and $vrs$ back to $U$.
  \end{packed_enum}
  \citem User $U$ encrypts $shares$ to $onions$ with public keys of selected trustees.
  \item User $U$ takes keccak256 hash $h(onion)$ of each $onion$ and submits the hash values to $C_s$.
\end{packed_enum}

\noindent \textbf{Second-round trustee selection:} 
\begin{packed_enum}[leftmargin=*]
  \setcounter{enumi}{9}
  \citem User $U$ randomly selects $n$ trustees and sends each trustee a $(sid,tid,onion)$, where $tid$ is non-repeated in $[n(l-1),nl)$.
  \citem Each selected trustee $T$ follows step 7, but in addition verifying received $onion$ with $h(onion)$ in $C_s$.

\end{packed_enum}
\noindent \textbf{Ciphertext and hash disclosure:} 
\begin{packed_enum}[leftmargin=*]
	\setcounter{enumi}{11}
	\citem User $U$ encrypts $(IN, vrs, R_U)$ with $key$ and make $E(key, (IN, vrs, R_U))$ public.
    \item User $U$ submits keccak256 hash $h(IN,R_U)$ and each trustee's $h(T^{addr},R_T)$ to $C_s$.
\end{packed_enum}

\end{mdframed}
\end{small}


\noindent \textbf{Step 2}:
The total remuneration that should be paid by user $U$ is $r=nlr_t+r_e$, where $r_d$ is a fixed per trustee remuneration hard-coded in $C_s$ and $r_e$ is a fixed amount of reward hard-coded in $C_m$ paying to the first trustee calling $execute()$ in $C_t$ during $w_e$. Both $r_d$ and $r_e$ can only be withdrawn by trustees after the end of execution time window $w_e$. 


\noindent \textbf{Step 4}:
After the schedule has been registered in $C_s$, the on-chain schedule information cannot be modified. Therefore, the information can be used by trustee candidates later in step 7 and 11 to verify the information transmitted through off-chain whisper channels from user $U$.

\noindent \textbf{Step 5}:
The Shamir secret sharing scheme~\cite{shamir1979share} with parameter $(m,n)$ can split the $key$ to $n$ $shares$. Later, any $m$ $shares$ among the $n$ can be combined to restore the $key$ while even $m-1$ $shares$ fail to do it. 
Therefore, even if some $shares$ are compromised, the compromised $shares$ may be insufficient to restore the $key$ before execution window $w_e$ while the rest $shares$ may still be sufficient to restore the $key$ during $w_e$.
In the example of Figure~\ref{f1}, we set $(m,n)=(2,3)$, so three $shares$ are generated from $key$ after splitting.

\noindent \textbf{Step 6-13}:
The design of two-round trustee selection implements the decentralized secret trust. The trustees selected in the first round should agree the user encrypt the $shares$ with their public keys for multiple layers so that the $shares$ become $onions$~\cite{dingledine2004torr} and harder to be compromised. Then, the trustees selected in the second round should take charge of storing these $onions$. Later, during $w_e$, once both the private keys of the first-round trustees and $onions$ stored by the second-round trustees are made public, the $key$ can be restored to decrypt the function inputs. 
The process offers following additional security features:

\begin{figure}
\centering
{
    \includegraphics[width=8.3cm,height=6cm]{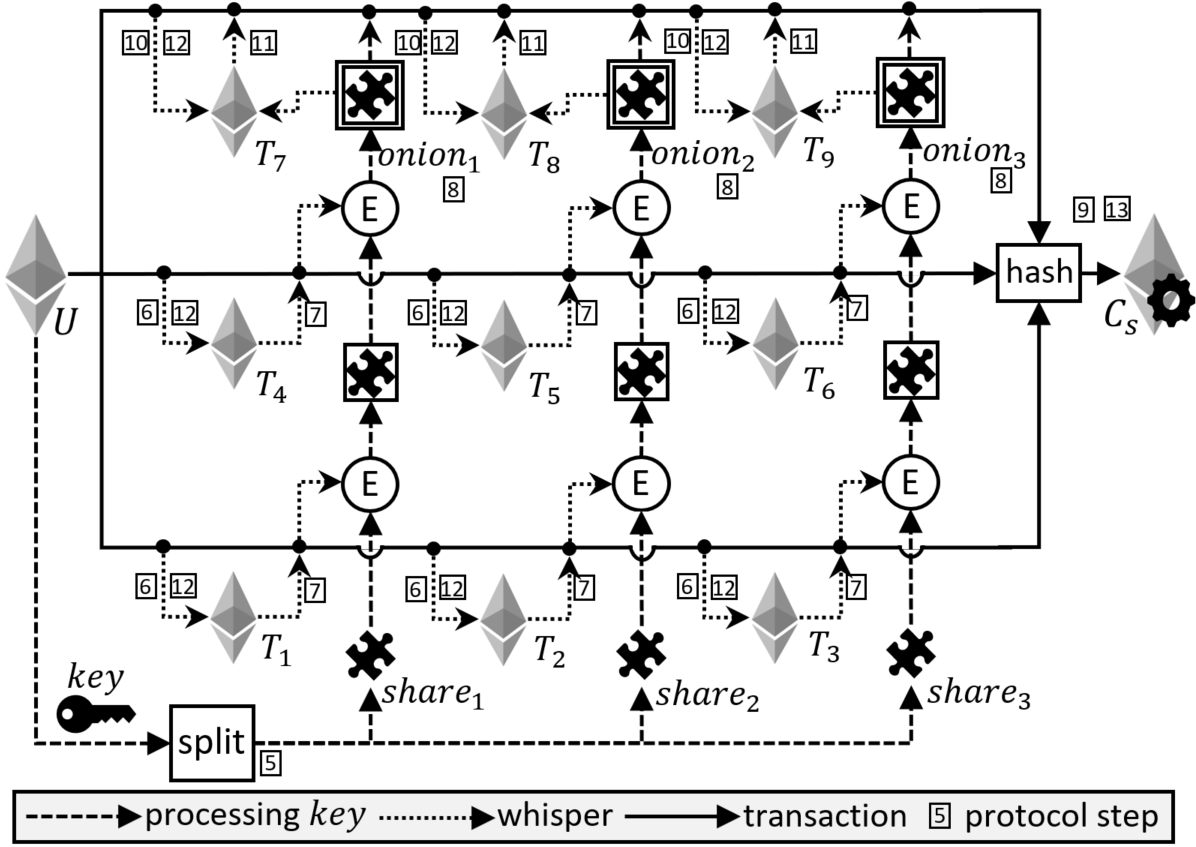}
}
\caption {User schedule example}
\label{f2} 
\end{figure}

\begin{itemize}[leftmargin=*]

\item The identities of selected trustees are kept private. 
In these steps, each trustee only communicates with the user through a whisper channel and all information that needs to be made public are announced by the user (step 9,12,13). Therefore, the identity of each trustee is only known to the user. This feature helps in suppressing collusion among trustees.

\item The identities of selected trustees are verifiable and only the trustees can pass the verification. 
To be verified as a specific trustee, both the trustee's address $T^{addr}$ and the nonce $R_T$ need to be submitted to $C_s$ and their hash should match with the one submitted by user in step 13. Since $R_T$ is created by the trustee, only the trustee has the ability to pass the verification. This feature also helps in suppressing collusion among trustees. We will discuss it in detail later in \textit{misbehavior report} component.

\item The identities of selected trustees are undeniable.
The user has signatures of the trustees (step 7,11) and the encrypted signatures are made public in step 12. Therefore, once $key$ is restored during $w_e$, the decrypted signatures can reveal the identities of all trustees.
This feature helps in detecting absent trustees who disappear during $w_e$.

\item The trustees are also protected against adversaries. 
It may be insecure to only allow users to publicly speak. Such a user may fabricate information and make trustees lose security deposit. To protect trustees from such users. Once a user has registered a schedule in step 4, the submitted information cannot be changed. Then, in step 7 and 11, each trustee can check the information before sending a signature to the user. 
This is also the main reason that we need two rounds. In step 11, the second-round trustees should first verify the correctness of the onions with the hash submitted by the user in step 9 and then provide signatures.

\end{itemize}

In the example of Figure~\ref{f2}, six trustees ($T_1$-$T_6$) are selected by user $U$ in the first round and their six public keys encrypt each of the three $shares$ with two layers, thus turning the $shares$ into two-layer $onions$. 
Then, three trustees ($T_7$-$T_9$) are selected by user $U$ in the second round to store the three $onions$.
Finally, $U$ ends the schedule by making the ciphertext public and submitting all hash values to 
$C_s$.

\subsection{Function Execution}
The third component of the protocol, \textit{function execution} indicates how the trustees selected in \textit{user schedule} component should collaboratively reveal the function inputs during execution window $w_e$ and send a transaction with the function inputs to the proxy contract $C_p$ through two phases, namely \textit{submission} (step 1-2) and \textit{execution} (step 3-6).


\begin{small}
\begin{mdframed}[innerleftmargin=8pt]
\centerline{\textbf{Function Execution}}
\textbf{Input:} scheduler contract $C_s$

\noindent \textbf{Submission (first half of $w_e$):} 
\begin{packed_enum}[leftmargin=*]
  \item Each trustee $T$ verifies its identity with $h(T^{addr},R_T)$ by submitting $R_T$ to $C_s$.
  \item Each trustee $T$ submits $onion$ or its private key to $C_s$, where $onion$ should be verified with $h(onion)$.
\end{packed_enum}

\noindent \textbf{Execution (second half of $w_e$):} 
\begin{packed_enum}[leftmargin=*]
	\setcounter{enumi}{2}
  \citem Any trustee $T$ can get $shares$ by decrypting $onions$ with the private keys.
  \citem Any trustee $T$ can get $key$ by combing any $m$ $shares$.
  \citem Any trustee $T$ can get $(IN, vrs, R_U)$ by doing $D(key,E(key, (IN, vrs, R_U)))$.
  \item  Any trustee $T$ can submit $(IN,R_U)$ to proxy contract $C_p$, where $(IN,R_U)$ can be verified with $h(IN,R_U)$ in $C_s$ and the correct function inputs $IN$ will trigger $C_p$ to call the target contract $C_t$.
\end{packed_enum}

\end{mdframed}
\end{small}

\noindent \textbf{Step 1-2}:
The \textit{submission} phase indicates the first half of execution window $w_e$, during which the protocol requires first-round and second-round trustees to submit their private keys and stored $onions$, respectively. To submit either a private key or an $onion$, a trustee should also provide the nonce $R_T$ generated in step 7 and 11 of \textit{user schedule} so that its identity can be verified with $h(T^{addr},R_T)$. 

\noindent \textbf{Step 3-6}:
The \textit{execution} phase refers to the second half of execution window $w_e$.
Since both $onions$ and private keys have been submitted, during this phase, any verified trustee should be able to turn $onions$ back to $shares$. 
Then, based on Shamir secret sharing scheme, any $m$ shares can be combined to restore the $key$ created by user $S$ in step 1 of \textit{user schedule}.
After getting the $key$, any trustee is able to decrypt the encrypted $(IN,vrs,R_U)$.
Finally, before the end of $w_e$, a verified trustee, after obtaining function inputs $IN$ and nonce $R_U$, should send proxy contract $C_p$ a transaction with both $IN$ and $R_U$.
Then, $C_p$ will immediately verify received $IN$ and $R_U$ with $h(IN,R_U)$ in scheduler contract $C_s$.
If both of them are correct, $C_p$ immediately send a message with $IN$ to the target contract $C_t$ to call the scheduled function.

\subsection{Misbehavior report}
The \textit{misbehavior report} represents the final component of the protocol and involves four types of misbehaviors that will result in the violator's security deposit being confiscated. All these misbehaviors are witnessable and the protocol rewards the reporter of a misbehavior a component of the violator's security deposit as an incentive while sending the rest of the violator's security deposit to the user.


\begin{small}
\begin{mdframed}[innerleftmargin=8pt]
\centerline{\textbf{Misbehavior report}}
\textbf{Input:} scheduler contract $C_s$

\noindent \textbf{Trustee identity disclosure report:} 
\begin{packed_enum}[leftmargin=*]
  \item Before the start of execution time $w_e$, any EOA can report a trustee identity disclosure misbehavior by submitting the nonce $R_T$ of the violator to scheduler contract $C_s$.
  \item If $h(T^{addr},R_T)$ using the submitted $R_T$  is same as the one in $C_s$, the misbehavior is verified.
\end{packed_enum}

\noindent \textbf{Advance disclosure report:} 
\begin{packed_enum}[leftmargin=*]\setcounter{enumi}{2}
  \item Before the start of execution time $w_e$, any EOA can report an advance disclosure misbehavior by submitting the private key belonging to the violator to scheduler contract $C_s$.
  \item If the public key derived from that private key is same as the violator's public key in $C_s$, the misbehavior is verified.
\end{packed_enum}

\noindent \textbf{Absent trustee report:} 
\begin{packed_enum}[leftmargin=*]\setcounter{enumi}{4}
  \item After step 5 in \textit{function execution}, any trustee can report an absent trustee misbehavior to scheduler contract $C_s$ by submitting the signature $vrs$ of the absent trustee.
  \item The address of the violator can be derived through $T = sigVerify((U^{addr},sid,tid,h(T^{addr},R_T)),vrs)$.
\end{packed_enum}

\noindent \textbf{Fake submission report:} 
\begin{packed_enum}[leftmargin=*]\setcounter{enumi}{6}
  \item After step 2 in \textit{function execution}, any trustee can report a fake submission misbehavior to scheduler contract $C_s$ if the trustee finds a submitted private key is incorrect.
  \item If the public key derived from that private key is different from violator's public key in $C_s$, the misbehavior is verified.
\end{packed_enum}

\end{mdframed}
\end{small}

\noindent \textbf{Trustee identity disclosure report}:
This report mechanism is designed to handle the trustee identity disclosure misbehavior presented in Section~\ref{s2.4}.
Before the start of execution window $w_e$, a trustee may choose to reveal its identity to seek collusion. 
To prove its identity, the violator has to reveal the nonce $R_T$ created by itself in step 7/11 of \textit{user schedule} so that its identity can become verifiable through $h(T^{addr},R_T)$ in $C_s$.
However, with this report mechanism, any EOA, after knowing $R_T$ before $w_e$, can report it to $C_s$ to earn reward.

\noindent \textbf{Advance disclosure report}:
The advance disclosure misbehavior introduced in Section~\ref{s2.4} can be handled using this report mechanism.
Before the start of $w_e$, a round-one trustee may choose to disclose its private key, which may help an adversary to decrypt $onions$ to $shares$, restore $key$ and obtain $IN$ before the start of $w_e$. 
However, with this report mechanism, any EOA, after knowing violator's private key before $w_e$, can betray the violator by reporting it to $C_s$.

\noindent \textbf{Absent trustee report}:
This report mechanism handles the absent trustee misbehavior described in Section~\ref{s2.4}.
Any trustee may become absent during $w_e$, thus increasing the failure chance of schedule.
With this report mechanism, any trustee, after obtaining signatures of all other trustees in step 5 of \textit{function execution}, can locally verify attendance of all other trustees, thus being able to report absent trustees to $C_s$.

\noindent \textbf{Fake submission report}:
Finally, the design of fake submission report aims at dealing with the fake submission misbehavior presented in Section~\ref{s2.4}.
In step 2 of \textit{function execution}, a submitted private key may not be the right one. 
Any trustee can locally verify a private key submitted by a suspect trustee through deriving the corresponding public key from the private key and comparing it with the public key submitted by that suspect trustee during \textit{trustee application}, thus becoming able to report violators to $C_s$.

\section{Security analysis}
\label{s4}
Next, we analyze the security guarantees of the proposed approach based on the rational adversary model. 
Recently, it has been widely recognized that assuming an adversary to be semi-honest or malicious is either too weak or too strong and hence modeling adversaries with rationality is a relevant choice in several attack scenarios~\cite{dong2017betrayal,nguyen2013analyzing}. Informally, a semi-honest adversary follows the prescribed protocol but tries to glean more information from available intermediate results while a malicious adversary can take any action for launching attacks~\cite{hazay2010note}. A rational adversary lies in the middle of the two types. 
That is, rational adversaries are self-interest-driven, they choose to violate protocols, such as colluding with other parties, only when doing so brings them a higher profit. 
In this paper, in order to design our approach with strong and practical security guarantees, we model all EOAs to be rational adversaries without assuming any of them to be honest or semi-honest. 

\begin{figure}
\centering
\subfigure[{\small $n=5$}]
{
   \label{hipc_01}
   \includegraphics[width=0.42\columnwidth]{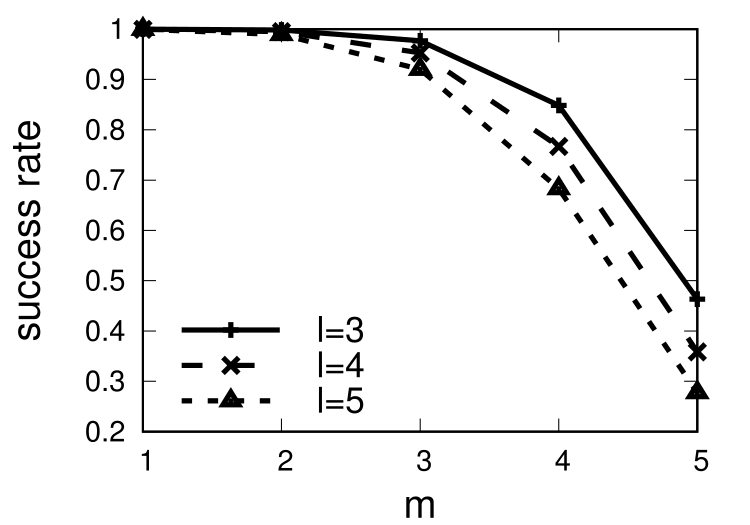}
}
\subfigure[{\small $n=10$}]
{
	\label{hipc_02}
    \includegraphics[width=0.42\columnwidth]{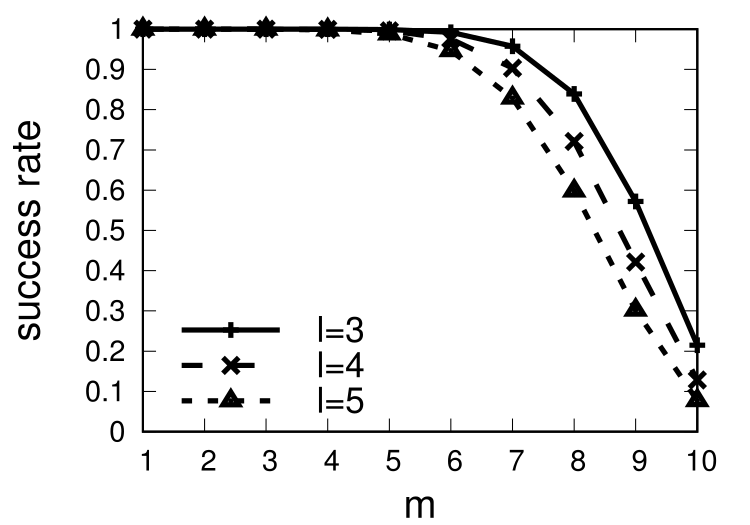}
 }
\caption{Schedule success rate when 5\% of trustees perform misbehaviors inadvertently}
\label {hipc_0102}
\end{figure}
Without countermeasures, such rational adversaries, after being selected by user as trustees, may perform four types of misbehaviors introduced in Section~\ref{s2.3}, including \textit{trustee identity disclosure}, \textit{advance disclosure}, \textit{absent trustee} and \textit{fake submission}.
As per the four misbehavior report mechanisms designed in \textit{misbehavior report} component, as long as the $key$ can be restored during the execution time window $w_e$, any of the four types of misbehaviors will lead to confiscation of the violator's deposit.
To prevent restoration of the $key$ so that misbehaviors can be performed in free, a certain fraction of trustees must collude to not submit their stored data (i.e., $onion$ or private key) together. However, due to trustee identity disclosure report mechanism in \textit{misbehavior report}, revealing trustee identity to other EOAs means losing deposit, so such a collusion will not happen among rational adversaries.

It is possible that a rational adversary performs misbehaviors inadvertently, such as forgetting providing the service or losing EOA's private key. 
Such kinds of inadvertent misbehaviors lead to same results of intentionally performing \textit{absent trustee} misbehavior. 
If we denote the percentage of EOAs performing inadvertent misbehaviors as $p_{IM}$, the success rate of a schedule with parameters $(l,m,n)$ will be computed through the Cumulative Distribution Function of Binomial distribution, namely $SR=1-\sum_{i=n-m+1}^{n} \binom{n}{i} P^i (1-P)^{n-i}$, where $P=1-(1-p_{IM})^l$ represents the probability that one $share$ is lost.
In Figure~\ref{hipc_0102}, 
we present the computed schedule success rate when 5\% of trustees perform misbehaviors inadvertently.
Specifically, in Figure~\ref{hipc_01}, by fixing $n$ to 5 and changing $m$ from 1 to 5, it shows that a smaller $m$, namely lower threshold for restoring $key$, performs higher resistance against inadvertent misbehaviors. By further changing $l$ from 3 to 5, we can find that a smaller $l$ offers better resistance against inadvertent misbehaviors. Then, in Figure~\ref{hipc_02}, $n$ is increased to 10. The increment of $n$ enhances the resistance against inadvertent misbehaviors when $m$ and $l$ do not change. Thus, larger $l$ and $n$ while smaller $m$ help maintaining high resistance against inadvertent misbehaviors.

\section{Implementation} 
\label{s5}
In this section, we present the implementation of the proposed protocol and discuss the experimental evaluation of the proposed mechanism in Ethereum.

\subsection{Implementation of protocol}
We first introduce the implementation setup and then present both key off-chain functions in node.js and on-chain functions in \textit{Solidity}~\cite{Solidity2017} and demonstrate how they work in practice. After that, we present two test instances used in our experimental evaluation.

\noindent \textbf{Setup}:
We programmed the smart contracts in \textit{Solidity}~\cite{Solidity2017}, the most commonly used smart contract programming language, deployed them to the Ethereum official test network \textit{rinkeby}~\cite{Rinkeby2017} and tested them with Ethereum official Go implementation \textit{Geth}~\cite{Geth2017}. 
Our experiments are performed on an Intel Core i7 2.70GHz PC with 16GB RAM.

\begin{table}
\centering
\begin{tabular}{|c |c |p{1.5cm} |p{3.5cm}|} \toprule 
    {\textbf{Component}} & {\textbf{Step}} & {\textbf{Function}} & {\textbf{Purpose}} \\ \midrule
    \multirow{4}{*}{\textbf{Schedule}}
    & 5 & share   & split $key$ to $shares$ \\
    & 7,11 & ecsign  & sign data with private key \\
    & 8 & encrypt  & encrypt $shares$ to $onions$ \\
    & 9,13 & soliditySha3  & compute keccak256 hash \\
    \midrule
    \multirow{2}{*}{\textbf{Execute}}
    & 3 & combine  & combine $shares$ to $key$ \\
    & 4 & decrypt  & decrypt $onions$ to $shares$ \\
    \bottomrule
\end{tabular}
\caption{Key off-chain functions in node.js, \textit{share()} and \textit{combine()} are in secrets.js~\cite{secrets.js}, \textit{ecsign()} is in ethereumjs-util~\cite{ethereumjs-util}, \textit{encrypt()} and \textit{decrypt()} are in eth-ecies~\cite{eth-ecies}, \textit{soliditySha3()} is in web3-utils~\cite{web3-utils}}
\label{t1}
\end{table}

\begin{table}
\centering
\begin{tabular}{|c |c |p{1.5cm} |p{3.2cm}|} \toprule 
    {\textbf{Component}} & {\textbf{Step}} & {\textbf{Function}} & {\textbf{Purpose}} \\ \midrule
    \textbf{Apply} & 2,3 & newCandidate & join candidate pool \\ \midrule
    \multirow{4}{*}{\textbf{Schedule}}
    & 4 & newUser   & register as a new user\\
    & 4 & newSchedule  & initialize a new schedule \\
    & 9 & setOnion  & submit hashes of onions \\
    & 13 & setTrustee  & submit hashes of trustees \\ \midrule
    \multirow{5}{*}{\textbf{Execute}}
    & \cellcolor{gray!25}1,2 & \cellcolor{gray!25}submitPrivkey  & \cellcolor{gray!25}submit private key \\ 
    & \cellcolor{gray!25}1,2 & \cellcolor{gray!25}submitOnion  & \cellcolor{gray!25}submit onion \\
    & \cellcolor{gray!25}6 & \cellcolor{gray!25}execute  & \cellcolor{gray!25}execute the target contract \\ 
    & 7 & withdrawD   & withdraw security deposit \\
    & 7 & withdrawR  & withdraw remuneration \\ \midrule
    \multirow{6}{*}{\textbf{Report}}
    & 1,2 & identityReport  & report identity disclosure \\
    & 3,4 & advanceReport  & report advance disclosure \\
    & 5,6 & absentReport  & report absent trustee \\
    & 7,8 & fakeReport  & report fake submission \\
    & 2,4,6,8 & withdrawA  & withdraw report award \\ \bottomrule
\end{tabular}
\caption{Key on-chain functions in solidity, the three colored functions are in proxy contract $C_p$, the rest of the functions are in scheduler contract $C_s$}
\label{t2}
\end{table}

\noindent \textbf{Implemented functions}:
The protocol primarily relies on 6 off-chain functions shown in Table~\ref{t1} and 15 on-chain functions shown in Table~\ref{t2}.  
In both the tables, we show the components and steps where each function works in protocol.
For example, function $share()$ is used in step 5 of \textit{user schedule} component to split $key$ to $n$ $shares$ using Shamir secret sharing~\cite{shamir1979share}.

\begin{itemize}[leftmargin=*]
\item \textbf{\textit{Trustee application}}: 
Any EOA in the network can invoke \textit{newCandidate()} to join the trustee candidate pool maintained by scheduler contract $C_s$.
\item \textbf{\textit{User schedule}}: 
Any EOA can invoke \textit{newUser()} to be recorded as a user and then set up new schedule through \textit{newSchedule()}. Then, during whisper communication with trustees, $h(onion)$ should be submitted to $C_s$ through \textit{setOnion()} while $h(T^{addr},R_T)$ and $h(IN,R_U)$ should be submitted to $C_s$ through \textit{setTrustee()}. 
Meanwhile, the generation of $shares$, signatures, $onions$ and hash values are completed by \textit{share()}, \textit{ecsign()}, \textit{encrypt()}, \textit{soliditySha3()} in node.js, respectively.
\item \textbf{\textit{Function execution}}: 
A trustee can submit private key and $onion$ through \textit{submitPrivkey()} and \textit{submitOnion()}, respectively. Then, after decrypting $onions$ to $shares$ through \textit{decrypt()} and combining $shares$ to $key$ through \textit{combine()}, any trustee has the ability to make the target function be executed through \textit{execute()}. Finally, after the execution window is over, trustees can withdraw deposit and remuneration through \textit{withdrawD()} and \textit{withdrawR()}, respectively.
\item \textbf{\textit{Misbehavior report}}: 
The four types of report mechanisms are implemented by \textit{identityReport()}, \textit{advanceReport()}, \textit{absentReport()} and \textit{fakeReport()}, respectively. Then, after the execution window is over, reporters can withdraw reward through \textit{withdrawA()}.
\end{itemize}

\begin{figure}
\centering
\subfigure[{\small Instance A}]
{
   \label{hipc_05}
   \includegraphics[width=0.42\columnwidth]{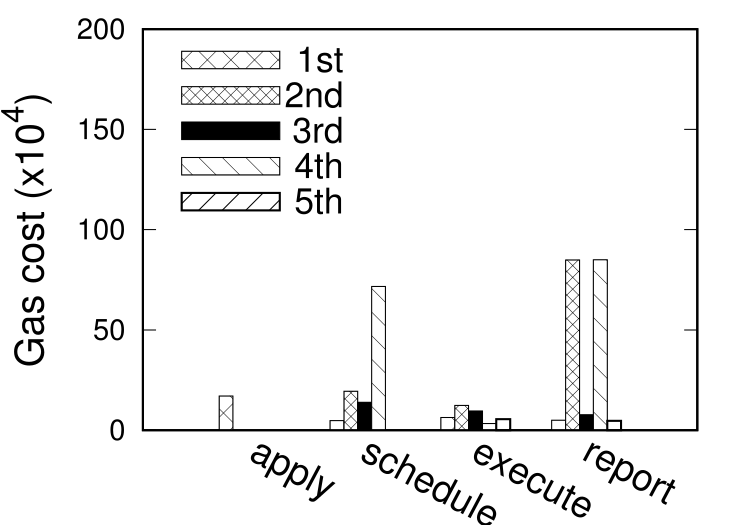}
}
\subfigure[{\small Instance B}]
{
	\label{hipc_06}
    \includegraphics[width=0.42\columnwidth]{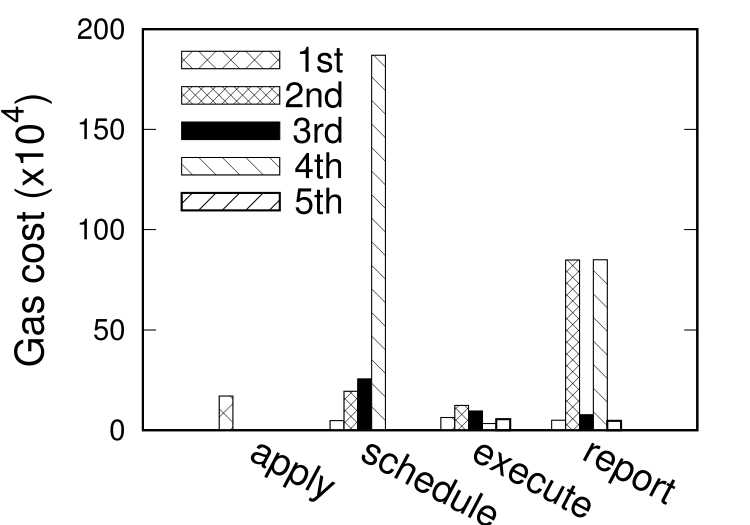}
 }
\caption{Gas cost}
\label {hipc_0506}
\end{figure}

\eat{
\noindent \textbf{Test instance}:
We design two test instances A and B:
\begin{table}[h]
\centering
\begin{tabular}{|c |c |c |c |c |c |c |} \toprule 
{\textbf{Instance}} & {\textbf{l,m,n}} & {\textbf{5\% IM}} & {\textbf{50\% M}} & {\textbf{L1}} & {\textbf{L2}} & {\textbf{L3}} \\ \midrule
    A & 3,2,5  & 99.82\% & 98.44\% & 4d & 4d  & vd \\
    \midrule
    B & 4,4,10 & 99.95\% & 99.99\% & 7d & 12d & 2vd \\
    \bottomrule
\end{tabular}
\label{t3}
\end{table}
}

\noindent \textbf{Test instance}:
We design two test instances A and B:
\begin{table}[h]
\centering
\begin{tabular}{|c |c |c |c |c |} \toprule 
{\textbf{Instance}} & {\textbf{l}} & {\textbf{m}} & {\textbf{n}} &  {\textbf{5\% IM}} \\ \midrule
    A & 3 & 2 & 5  & 99.82\% \\
    \midrule
    B & 4 & 4 & 10 & 99.95\%  \\
    \bottomrule
\end{tabular}
\label{t3}
\end{table}

Instance A employs 15 trustees while instance B employs 40 trustees. 
As a result, instance B has higher schedule success
rate under 5\% inadvertent misbehaviors (IM).
In both instance A and B, we use the \textit{SealedBidAuction} contract~\cite{SealedBidAuction} as the target contract $C_t$ and we assumed user's goal was to schedule a transaction calling function $reveal(amount,nonce)$.
Specifically, we designed an input parameter $time$ to simulate the time during testing.


\subsection{Experimental evaluation} 

We use the presented test instances to experimentally evaluate the performance of the smart contracts, namely the gas cost and time overhead of each function presented in Table~\ref{t2}.

\noindent \textbf{Gas cost}:
Gas is spent in Ethereum for deploying smart contracts or calling functions.
The gas costs of functions in Table~\ref{t2} for instance A and B are shown in Figure~\ref{hipc_05} and Figure~\ref{hipc_06}, respectively. 
For ease of presentation, results are grouped into four clusters. Each cluster represents a protocol component and contains a group of functions following their order in Table~\ref{t2}.
As can be seen, most functions cost very little. Specifically, among the fifteen functions, eight cost lower than $10^5$ gas and eleven cost lower than $2 \times 10^5$ gas. 
Among the rest four functions, both \textit{advanceReport()} and \textit{fakeReport()} cost around $8.5 \times 10^5$ because the two functions need to derive public key from private key on chain.
Gas costs of the last two functions, namely $setOnion()$ and $setTrustee()$, change with $n$ and $nl$, respectively.
From instance A to B, $l$ increases from $3$ to $4$ and $n$ increases from $5$ to $10$. As a result, gas cost of $setOnion()$ increases from $1.40 \times 10^5$ to $2.55 \times 10^5$ and gas cost of $settrustee()$ increases from $7.17 \times 10^5$ to $1.87 \times 10^6$.

To complete a schedule, some functions need to be invoked for multiple times. Below, we show the number of times that each function needs to be invoked in a single schedule when there is no report needed:

\begin{table}[h]
\centering
\begin{tabular}{|c c|c c|c c|} \toprule 
{\textbf{Function}} & {\textbf{No.}} & {\textbf{Function}} & {\textbf{No.}} & {\textbf{Function}} & {\textbf{No.}} \\ \midrule
    newCandidate & $nl$ & setTrustee  & $1$ & execute & $1$  \\
    newSschedule   & 1  & submitPrivkey & $n(l-1)$ & withdrawD & $nl$  \\
    setOnion     & $1$  & submitOnion   & $n$ & withdrawR & $nl$  \\ 
    \bottomrule
\end{tabular}
\label{t4}
\end{table}

\begin{figure}
\centering
\subfigure[{\small Instance A}]
{
   \label{hipc_07}
   \includegraphics[width=0.42\columnwidth]{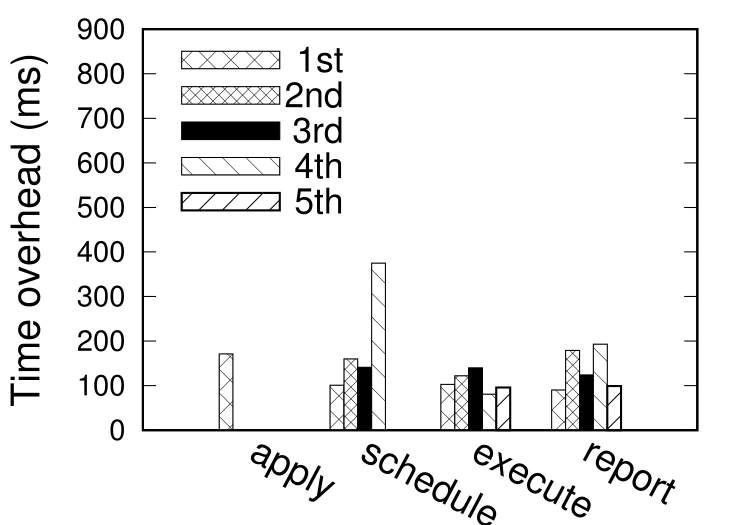}
}
\subfigure[{\small Instance B}]
{
	\label{hipc_08}
    \includegraphics[width=0.42\columnwidth]{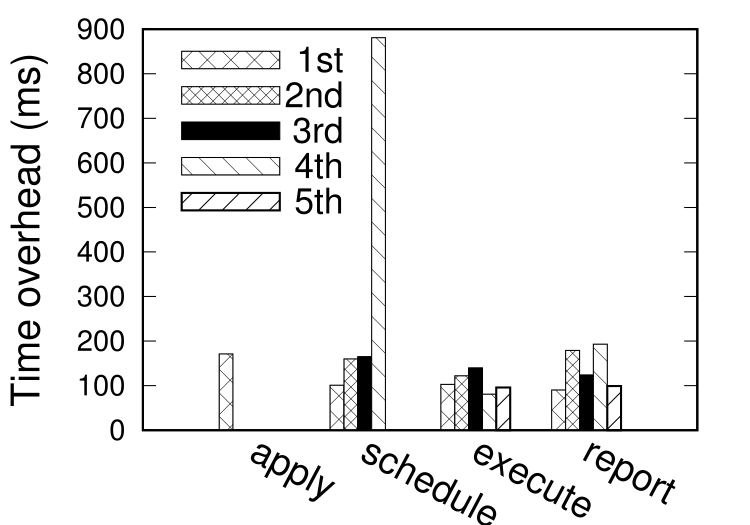}
 }

\caption{Time overhead}
\label {hipc_0708}

\end{figure}

Besides, the gas cost of deploying proxy contract $C_p$ is about $1.33 \times 10^6$. Therefore, the total gas costs of instance A and B are $7.60 \times 10^6$ and $1.72 \times 10^7$, respectively.
Both gas price and Ether price keeps dramatically swinging~\cite{etherscan}. For example, based on prices of date 12/5/2016, instance A and B cost \$1.2 and \$2.72, respectively. However, based on prices of date 10/29/2017, the two instances cost \$22.8 and \$51.6, respectively. As can be seen, the monetary cost of a timed-execution service is highly influenced by the fluctuation of cryptocurrency market, which may be a common limitation of cryptocurrency-based applications.

\noindent \textbf{Time overhead}:
The time overheads of functions in Table~\ref{t2} for instance A and B are shown in Figure~\ref{hipc_07} and Figure~\ref{hipc_08}, respectively. 
All results are averaged for 100 tests.
Among the fifteen functions, fourteen functions spend 0-200ms.
It is the function \textit{setTrustee()} that spends more time to record information of all the trustees to the blockchain. Specifically, \textit{setTrustee()} spends 375ms for instance A while 881ms for instance B as there are more trustees in instance B.

\section{Related work}
\label{s6}
The problem of revealing private data at a release time in future has been researched for more than two decades. The problem was first described by May as timed-release cryptography in 1992~\cite{may1992timed} and has intrigued many researchers since then. 
There are four sets of representative solutions in the literature.
The first category of solutions was designed to make data recipients solve a mathematical puzzle, called time-lock puzzle, before reading the messages~\cite{bitansky2016time,boneh2000timed,rivest1996time}.
The time-lock puzzle can only be solved with sequential operations, thus making multiple computers no better than a single computer. This solution suffers from two key drawbacks. First, the time taken to solve a puzzle may be different on different computers. Second, the puzzle computation is associated with a significant computation cost, which does not lead to a scalable cost-effective solution.
The second group of solutions relies on a third party, also known as a time server, to release the protected information at the release time in future. The information, sometimes called time trapdoors, can be used by recipients to decrypt the encrypted message~\cite{kasamatsu2012time,rivest1996time}. 
However, the time server in this model has to be trusted to not collude with recipients so that encrypted messages cannot be entered before release time. This restriction makes this set of solutions involve a single point of trust.
The third set of approaches studied the problem in the context of Distributed Hash Table (DHT) networks~\cite{li2017emerge,li2017timed}. The idea behind these techniques is to leverage the scalability and distributed features of DHT P2P networks to make message securely hidden before release time.
Finally, the last direction uses blockchains as a reference time clock correctness guaranteed by the distributed network~\cite{jager2015build,liu2015time}. By combining witness encryption~\cite{garg2013witness} with blockchain, one can leverage the computation power of PoW in blockchain to decrypt a message after a certain number of new blocks have been generated. However, the current implementation of witness encryption is far from practical, which requires an astronomical decryption time estimated to be $2^{100}$ seconds~\cite{liu2015time}. 
Recent work has studied the problem of supporting self-emerging data in blockchain networks~\cite{srds}. It allows the encrypted private data to travel through a long path within a blockchain network and appear at the prescribed release time.
However, unlike the proposed work in this paper, this approach supports only self-emergence of data and fails to support timed invocation of smart contract functions which is crucial for supporting timed executions in decentralized applications on blockchain-based smart contract platforms.
Besides the above mentioned solutions, there are two tools that support timed execution of transactions, however, they do not protect sensitive inputs. 
\textit{Ethereum Alarm Clock}~\cite{Clock} allows a client to deploy a request contract to the Ethereum network at time A with a reward and if any account is interested in the reward, the account can invoke the request contract at a prescribed Time B to make the scheduled transaction be sent to earn the reward. However, this scheme neither protects sensitive inputs nor guarantees the transaction execution.
\textit{Oraclize}~\cite{oraclize} is a blockchain oracle service that takes the role of a trusted third party (TTP) to execute the transaction on behalf of the client at a future time point. The limitations of of this scheme include both the centralization brought by the TTP and the lack of protection of sensitive inputs.
To the best of our knowledge, the approach proposed in this paper is the first decentralized solution for enabling users of decentralized applications to schedule timed execution of transactions without revealing sensitive inputs before an execution time window chosen by the users.

\section{Conclusion}
\label{s7}

In this paper, we developed a new decentralized privacy-preserving timed execution mechanism that allows users of Ethereum-based decentralized applications to schedule timed transactions without revealing sensitive inputs before an execution time window chosen by the users.
The proposed approach involves no centralized party and allows users to go offline at their discretion after scheduling a timed transaction.
The timed execution mechanism protects the sensitive inputs by employing a set of trustees from the decentralized blockchain network to enable the inputs to be revealed only during the execution time.
We implemented the proposed approach using \textit{Solidity} and evaluated the system on the Ethereum official test network.
Our theoretical analysis and extensive experiments validate the security properties and demonstrate the low gas cost and low time overhead associated with the proposed approach.

\renewcommand\refname{Reference}
\bibliographystyle{IEEEtran}
\bibliographystyle{plain}
\urlstyle{same}

\bibliography{main.bbl}

\end{document}